\documentclass[%
 reprint,
 amsmath,amssymb,
 aps,
tikz,
crop,
convert={density=400,outext=.png},
]{revtex4-2}

\usepackage{amsmath}

\usepackage{amsmath}
\usepackage{amssymb}
\usepackage{tikz}
\usepackage{quantikz}

\usepackage{subcaption}
\usepackage{caption}

\usepackage{siunitx}
\usepackage{xcolor}
\newcolumntype{R}{S[table-format=2.1,round-mode=places,round-precision = 3,color=blue,negative-color = red]}
\usepackage{booktabs}
\usepackage{sparklines}

\usepackage[labelfont=bf]{caption}

\usepackage{multirow}
\usepackage{array}
\usepackage{arydshln}
\def\sparkrectangleh #1 #2 {%
   \ifdim #1pt > #2pt
        \errmessage{The left corner #1 of rectangle cannot be lower than #2}%
   \fi
   {\pgfmoveto{\pgforigin}\color{sparkrectanglecolor}%
   \pgfrect[fill]{\pgfxy(#1, 0)}{\pgfxy(#2-#1,1)}}}%

\usepackage{array}{}
\usepackage[round-mode=places,detect-weight=true,detect-inline-weight=math]{siunitx}

\def\sparklineheight ex{7pt}

\usepackage{booktabs}
\usepackage{etoolbox}
\usepackage[round-mode=places,detect-weight=true,detect-inline-weight=math]{siunitx}

\usepackage{graphicx}
\usepackage{dcolumn}
\usepackage{bm}
\usepackage{braket}
\usepackage{comment}

\begin{document}

\preprint{APS/123-QED}


\title{Bohr’s complementarity principle tested on a real quantum computer via interferometer experiments}

 \author{Celia Álvarez Álvarez $^1$}
\author{Mariamo Mussa Juane $^2$}
 \email{mmussa@cesga.gal} 

\affiliation{
$^1$ Universidade de Santiago de Compostela, Santiago de Compostela, Spain 
}
\affiliation{
$^2$Centro de Supercomputación de Galicia (CESGA), Santiago de Compostela, Spain 
}

\date{\today}

\begin{abstract}
Bohr's Complementarity Principle is a core concept of quantum mechanics. In this article, an updated complementarity relation for the wave and ondulatory aspects of a quantum system is presented and discussed. Two interferometric experiments are implemented in one and two qubit circuits and executed on real hardware. The final state density matrices are reconstructed using quantum state tomography and the complementarity relation is tested via direct computation. Results of the executions are presented both graphically and with a mean squared error analysis for a better comprehension.

\end{abstract}

\maketitle


\section{INTRODUCTION.}


  Any observation of atomic phenomena implies an interaction of the quantum system with the observing agent that cannot be dismissed. In his lectures at the Como Conference and Fifth Solvay Congress (1927), Niels Bohr first introduced the physical model of Complementarity, which later evolved into what is now known as Bohr's Complementarity Principle (BCP). The BCP introduces the notion of complementarity by reflecting on how the nature of the quantum theory forces to contemplate certain quantum phenomena as complementary and therefore excluding in the description of a quantum system, with the root of this complementarity residing in the fact that the required experimental contexts for each quantum property to be observed are mutually exclusive. In other words, each compendium of complementary quantum properties cannot be observed simultaneously and therefore known, but according to quantum theory they're all equally real \cite{Bohr_1928_QuantumPostulate}. \\

The BCP fueled the birth of the Einstein-Podolsky-Rosen (EPR) paradox as a reply. For the authors, complementarity wasn't a feature of nature but rather a flaw of our description of it. Therefore quantum mechanics was deemed to be an incomplete theory \cite{EPR_1935}. The Bohr-Einstein debates culminated in the Copenhagen interpretation: Bohr's view of quantum mechanics prevailing as a part of the modern understanding of physics.\\

A first approach to a quantitative model for complementarity was carried out in 1979 by Wooters and Zurek \cite{WZurek_1979}. In their article, they develop a theoretical treatment of information in the double slit experiment.
In 1988 Greenberg and Yasin introduce the wave-particle relation \cite{GREENBERGER_1988}, which quantifies how partial measurements of the which-way information or predictability (P) of the path of a particle and of the visibility (V) of the interference pattern of a wave, balance out in order to fulfill the inequality

\begin{equation}\label{ineqPV}
    P^2+V^2\leq1.
\end{equation}


A relation like the one shown in equation (\ref{ineqPV}) reveals that complementary characteristics of a quantum system are not necessarily mutually exclusive, that partial manifestation of ondulatory and corpuscular natures of one quantum system can happen for simultaneously in a balanced way. The Biased Mach-Zehnder Interferometer (BMZI) and Partial Quantum Eraser (PQE) are valid configurations to investigate the wave-particle duality. Interferometric experiments of the sort allow for the BCP to manifest by forcing the system to modify its initial state by creating superposition and/or entanglement, and later on making the wave function collapse with measurement.\\ 

A full quantum simulation of the BMZI has already been implemented on an IBM quantum computer \cite{BMZI_2023} and the BCP has been verified on this experiment. Similarly, simulations of Quantum Eraser interferometric experiments have also been carried out on IBM's quantum computers \cite{S_Chrysosthemos_2023}.\\ 

In this article, the main goals are: (1) implementing the BMZI and the PQE experiments to discuss them in relation to the complementarity relations of the sort of equation (\ref{ineqPV}), and (2) presenting the results for the executions the real quantum computer available at CESGA, QMIO\cite{cacheiro2025qmio}. 

\section{METHODS.}

\subsection{Coherence, predictability and continuous complementarity.}



Quantum coherence provides a natural extension for the visibility and therefore used in this work to quantify the ondulatory aspect of a qudit. The l$_1$-norm quantum coherence is defined \cite{Baumgratz2014}

\begin{equation}
    C_{l_1}(\rho)=\sum_{j\neq k} |\rho_{jk}|.
\label{eq:C}
\end{equation}

The properties of density matrices (represented by $\rho$), as consequence of the state and measurement formulation in quantum mechanics, provide an upper limit for this quantisation of the ondulatory aspect as a function of the off-diagonal elements of the density matrix. A density matrix must satisfy $\rho\geq0$ (positive definite), this implies $|\rho_{jk}|^2\leq\rho_{jj}\rho_{kk}$ for all $j, k$. Then the following inequality can be derived:

\begin{equation}
    C_{l_1}(\rho)\leq\sum_{j\neq k}\sqrt{\rho_{jj}\rho_{kk}}\leq d-1.
\end{equation}

A complete derivation is not presented in this work and further details can be consulted in \cite{Basso_2020} and \cite{durr_2001}. Here $d$ would be the number of modes for a qudit (or the dimension of the Hilbert space used to express the state of said qudit, for a qubit $d=2$). On the other hand, if we identify the following quantity as the predictability

\begin{equation}
    P_{l_1}(\rho)=d-1-\sum_{j\neq k}\sqrt{\rho_{jj}\rho_{kk}},
\label{eq:P}
\end{equation}

 the desired quantum complementarity inequality can be derived:

\begin{equation}
    C_{l_1}(\rho)+P_{l_1}(\rho)\leq d-1.
\label{eq:C+P}
\end{equation}

\subsubsection{A note on incoherence and purity.}

For a particular basis $\{ \ket{i}\} _{i=1,...,d}$ of the d-dimensional Hilbert space ($\mathcal{H}$) that a quantum state is considered in, a diagonal density operator $\hat{\delta}$ (in this particular basis) is an \textit{incoherent} state. In other words

\begin{equation}
\hat{\delta} \in I \Leftrightarrow \hat{\delta}=\sum_{i=1}^{d}\delta_i\ket{i}\bra{i},
\end{equation}

with $c_i \in \mathbb{C}$ and $\ket{i} \in \mathcal{H}$.

Incoherent states must not be confused with \textit{pure} states, which consist of density matrices of the form

\begin{equation}
    \rho =\ket{\psi}\bra{\psi}.
\label{eq:pure}
\end{equation}

For a pure state one expects inequality (5) to turn into an equality. This is demonstrated straightforwardly following the definition of a pure state. One can write

\begin{equation}
    \ket{\psi}=\sum_{i=1}^d c_i\ket{i} \Rightarrow \rho=\sum_{i,j=1}^d c_ic^*_j\ket{i}\bra{j}.
\end{equation}

By plugging this explicit expression of $\rho$ into (2) and (4):

\begin{equation}
    C_{l_1}(\rho)=\sum_{i\neq j} |\rho_{ij}|=\sum_{i\neq j}|c_ic_j^*|=\sum_{i\neq j}|c_i||c_j|,
\end{equation}

\begin{equation}
\begin{aligned}
    P_{l_1}(\rho)=d-1-\sum_{j\neq k}\sqrt{\rho_{ii}\rho_{jj}}=\\
    =d-1-\sum_{j\neq k}\sqrt{|c_ic_i^*c_jc_j^*|}
    =d-1-\sum_{i\neq j}|c_i||c_j|;
    \end{aligned}   
\end{equation}

one sees the equality is now fulfilled:
$C_{l_1}(\rho)+P_{l_1}(\rho)= d-1.$

\subsubsection{A note on independence.}

The space of all $\mathbb{C}^{n×n}$ matrices is a vector space of dimension $n^2$. In this space, the set of diagonal matrices ($\mathcal{D}$) and the set of off‑diagonal matrices ($\mathcal{O}$) are complementary linear subspaces, and every matrix decomposes uniquely as a sum of a diagonal part plus an off‑diagonal part, or in terms of subspaces, $\mathbb{C}^{n×n}=\mathcal{D}\oplus\mathcal{O}$. An analogous decomposition can be done for density matrices.\\

As presented above, coherence and predictability each depend on complementary parts of the density matrix. This alone would lead one to believe both variables are independent.
Nevertheless, the positivity of the density matrix enforces the trade-off relation (5), which turns into an equality for pure states and therefore a dependence between $C_{l_1}$ and $P_{l_1}$ can be established.


\subsection{Experiments to test Bohr's Complementarity Principle.}

Interferometric experiments, as the ones implemented in this work, allow studying Bohr's Complementarity Principle, as the initial state of the system of the experiment is brought into superposition and later on forced to collapse via measurement. By the execution on a high number of qubits, all initialised to the same state, the density matrix can be reconstructed for the final state of the system after the interferometric experiment using quantum state tomography. Finally, the density matrix reconstruction will provide a straightforward way to compute the quantity $C_{l_1}(\rho)+P_{l_1}(\rho)$ (as well as both terms of the sum independently).\\

In this section we implement the BMZI and the PQE in one and two qubit circuits following \cite{updatedPCB_2024}, with a final goal testing inequality (\ref{eq:C+P}).\\

\subsubsection{Biased Mach-Zehnder Interferometer. 1 Qubit.}

\begin{figure}[!hb]
    \centering
\begin{quantikz}
    \lstick{$q$} 
    & \gate{R_X(-1.0\alpha)}
    & \gate{iX} 
    & \gate{P(\phi)} 
    & \gate{R_X(-1.0\beta)} 
    & \qw
\end{quantikz}
    \caption{Implementation of the BMZI on a 1-qubit quantum circuit with qiskit. The angles $\phi$ and $-\beta$ will be fixed to 0 and $\pi$, respectively.}
    \label{fig:bmzi}
\end{figure}
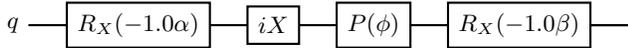

As a first step to implement the BMZI we simulate the optical elements from the setting provided in figure 1 of reference \cite{updatedPCB_2024} using unitary gates as follows:\\

\noindent(1) The states $\ket{0}$ and $\ket{1}$ codify the horizontal and vertical spatial modes, respectively, for the quantum system (i.e. photon path).\\
(2) The initial state being $\ket{\psi_0}=\ket{0}$, goes through the first biased beam splitter (BBS) implementing the application of the gate $R_x(-\alpha)$=$\begin{bmatrix} \text{cos}(\alpha/2) & i \text{ sin}(\alpha/2) \\ i \text{ sin}(\alpha/2) & \text{cos }(\alpha/2) \end{bmatrix}$.\\
(3) Then the combined action of the mirrors is implemented using $iX=\begin{bmatrix} 0 & i\\ i & 0 \end{bmatrix}$.\\
(4) A phase gate $P(\phi)=\begin{bmatrix} 1 & 0\\ 0 & e^{i\phi} \end{bmatrix}$ performs a phase shift.\\
(5) Finally, we apply another BBS through the gate $R_X(-\beta)$.\\

The final state is:

\begin{equation}
    \ket{\psi_3}=-(e^{i\phi}T_1R_2+R_1T_2)\ket{0}+i(e^{i\phi}T_1T_2-R_1R_2)\ket{1},
\end{equation}

where $T_1$, $T_2$, $R_1$, $R_2$ are the transmission and reflection coefficients of BBS1 and BBS2. These should fulfill $T^2+R^2=1$ and therefore can be parametrised by one angle ($T_1=cos(\alpha/2)$, $R_1=sin(\alpha/2)$, $T_2=cos(\beta/2)$ and $R_2=sin(\beta/2)$). It is straightforward to check that the final states fulfills $C_{l_1}(\rho)+P_{l_1}(\rho)=1$ \cite{updatedPCB_2024}. This is the ideal case. For any quantum computer we expect a lower value subjective to the noise of the device. \\

Theoretical values for $C_{l_1}(\rho)$ and $P_{l_1}(\rho)$ can be derived by explicit computation of the density matrix of the final state: $\rho=\ket{\psi_3}\bra{\psi_3}$.

\subsubsection{Partial Quantum Eraser. 2 Qubits.}

\begin{figure*}
    \centering
    \begin{quantikz}
        \lstick{$q_0$} 
        & \qw\slice[style=gray]{} 
        & \targ{}\slice[style=gray]{} 
        & \qw 
        & \qw \slice[style=gray]{} 
        & \qw \slice[style=gray]{} 
        & \gate{H}\slice[style=gray]{} 
        & \ctrl{1} 
        & \qw \\
        \lstick{$q_1$} 
        & \gate{R_X(-\frac{\pi}{2})} 
        & \ctrl{-1}  
        & \gate{iX} 
        & \gate{P(\phi)}
        & \gate{R_X(-\frac{\pi}{2})}  
        & \octrl{-1}
        & \gate{iX}
        & \qw
    \end{quantikz}
    \caption{Implementation of the PQE on a 2-qubit quantum circuit with qiskit.}
    \label{fig:pqe}
\end{figure*}
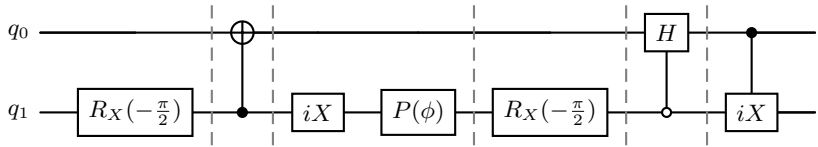

For the implementation of the PQE, two additional polarisation modes are defined for the system ($\ket{0}$ for horizontal (H) polarisation and $\ket{1}$ for vertical (V) polarisation. We simulate the optical elements from the setting provided in figure 2 of reference \cite{updatedPCB_2024}  as shown in FIG. \ref{fig:pqe}:\\

\noindent (1) The polarisation modes would be now encoded by adding a second qubit ($q_0$ in FIG. \ref{fig:pqe}) to our quantum circuit and system will be initially in the state $\ket{q_1q_0}=\ket{00}$.\\
(2) The initial state is altered by the action of the first unbiased BS $R_x(-\pi/2)$=$\begin{bmatrix} 1/\sqrt{2} & i/\sqrt{2}  \\ i/\sqrt{2} & 1/\sqrt{2} \end{bmatrix}$ on $q_1$.\\
(3) Then a half-wave plate (HWP) is implemented with a CX gate ($\ket{0}\bra{0}\otimes I+ \ket{1}\bra{1}\otimes X$), changing the polarisation from H to V and vice versa if the spatial mode corresponds to $\ket{1}$.\\
(4) The mirrors and phase shift act on $q_1$ and are implemented using $iX$ and $P(\phi)$ as in the BMZI.\\
(5) After, a second BS is applied ($R_x(-\pi/2)$) on $q_1$.\\
(6) Then a quarter-wave plate (QWP) is applied with a controlled circuit gate ($\ket{0}\bra{0}\otimes H+ \ket{1}\bra{1}\otimes I$), adding superposition on the polarization. \\
(7) Finally, polarizing beam splitters (PBS) are both added to the vertical and horizontal paths ($I\otimes\ket{0}\bra{0}+ iX\otimes\ket{1}\bra{1}$). The resulting state after the PQE would be

\begin{equation}
\begin{aligned}
    \ket{\psi_4}_{in} = -\frac{1}{2\sqrt{2}}\Big[
    &(e^{i\phi}+1)\ket{00}
      -(e^{i\phi}-1)\ket{11} \\
    &-i\sqrt{2}\,e^{i\phi}\ket{10}
      -\sqrt{2}\ket{01}
    \Big].
\end{aligned}
\end{equation}

Theoretical values for $C_{l_1}(\rho)$ and $P_{l_1}(\rho)$ can be derived by explicit computation of the density matrix of the final state: $\rho=\ket{\psi_4}_{in}\bra{\psi_4}_{in}$.

\subsection{Deconstructed mean squared error}

Both experiments, BMZI and PQE, allow measuring coherence and predictability for different angles to verify the inequality (\ref{eq:C+P}) for all the cases. We reconstruct the density matrix through a basic quantum state tomography (reference) to obtain $C_{l_1}$ and $P_{l_1}$ regarding equations (\ref{eq:C}) and (\ref{eq:P}). We aim to analyse the deviation of the sum of the obtained coherence and predictability regarding the theoretical value $1$ with the mean squared error (MSE). This sum  $\mathbf{C_{l_1}}+\mathbf{P_{l_1}}$ is an array with as many dimensions $n$ as set angles, $\alpha$ for the BMZI and $\phi$ for the PQE. As errors can be decomposed so can MSE \cite{hodson2021mean}. Therefore, the decomposition of the sum of the MSE is: 

\begin{equation}    
\begin{split}
  MSE(\mathbf{C_{l_1}}+\mathbf{P_{l_1}})=\\
  =\frac{1}{n}\sum_{i=1}^{n} \left(  \left(  \widehat{C}_{l_{1i}}-C_{l_{1i}} \right)  + \left(  \widehat{P}_{l_{1i}}-P_{l_{1i}} \right)\right)^2= \\ 
  = MSE (\mathbf{C_{l_1}}) + MSE (\mathbf{P_{l_1}}) + corr
\end{split}
\label{eq:MSE}
\end{equation}

where $X_{l_{1i}}$ is the experimental value and $\widehat{X}_{l_{1i}}$ the theoretical value associated to each angle of index $i=1,...,n$, and being $X=\{C,P\}$. The final term is analogous to a covariance between $\mathbf{C_{l_1}}$ and $\mathbf{P_{l_1}}$:

\begin{equation} 
corr = \frac{2}{n}\sum_{i=1}^{n}\left(  \widehat{C}_{l_{1i}}-C_{l_{1i}} \right)\left(  \widehat{P}_{l_{1i}}-P_{l_{1i}} \right)
\label{eq:corr}
\end{equation}

If $\mathbf{C_{l_1}}$ and $\mathbf{P_{l_1}}$ are highly correlated, the absolute value of the sum of the products is also high. The sum of the products cancels if there is no correlation. This is the extreme case of orthogonal variables where the MSE of the sum is equal to the sum of MSEs. 


\begin{figure*}[!ht]

\begin{minipage}[b]{0.45\textwidth}
   \centering
    \includegraphics[width=1\linewidth]{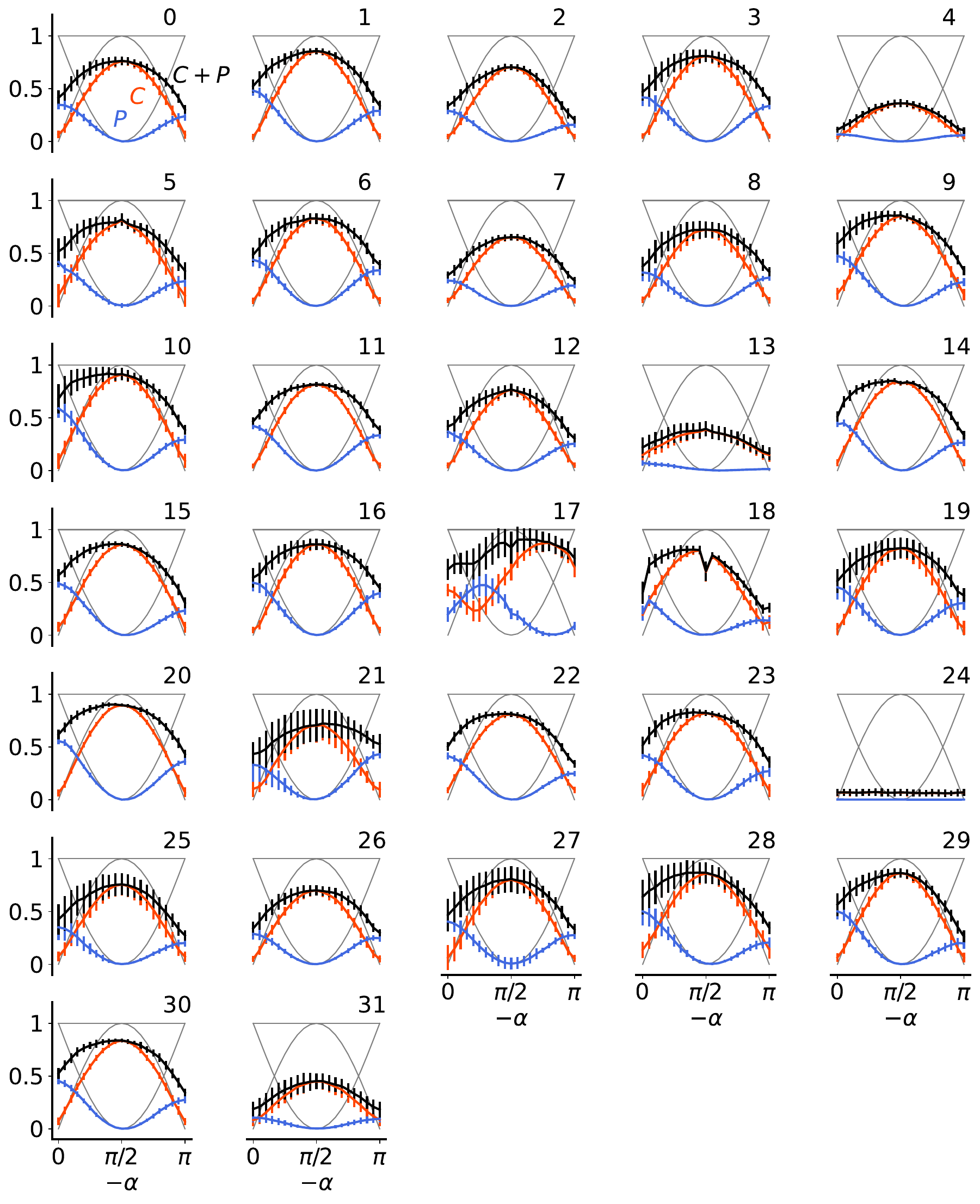}
      \bigskip
      \subcaption{BCP curves}  
      \label{fig:BMZI}
\end{minipage}  
\hspace{1cm}  
\begin{minipage}[b]{0.45\textwidth}

\robustify\bfseries
\tabcolsep=0.11cm
\scalebox{0.7}{
\begin{tabular}{lS[table-format=2.4,round-mode=places,round-precision = 3]S[table-format=2.4, round-mode=places, round-precision = 3]S[table-format=2.4,round-mode=places,round-precision = 3]S[table-format=2.4, round-mode=places,round-precision = 3]cS[table-format=2.3, round-mode=places,round-precision = 2]}
\toprule

\textbf{qubit ID} & \textbf{mean} & \textbf{std} & \textbf{corr} & \textbf{min} & \textbf{MSE histogram} & \textbf{max} \\
\midrule
0 & 0.163000 & 0.026000 & 0.017000 & 0.091000 & \begin{sparkline}{15}\sparkdot 0.2176139967112173 0 black \sparkdot 0.09094757043778409 0 black \spark0.0 0.0 0.0169 0.0 0.0339 0.0 0.0508 0.0 0.0678 0.0 0.0847 0.0968 0.1017 0.1613 0.1186 0.1935 0.1356 0.5484 0.1525 0.8065 0.1695 1.0 0.1864 0.4839 0.2034 0.2258 0.2203 0.0323 0.2373 0.0 0.2542 0.0 0.2712 0.0 0.2881 0.0 0.3051 0.0 0.322 0.0 0.339 0.0 0.3559 0.0 0.3729 0.0 0.3898 0.0 0.4068 0.0 0.4237 0.0 0.4407 0.0 0.4576 0.0 0.4746 0.0 0.4915 0.0 0.5085 0.0 0.5254 0.0 0.5424 0.0 0.5593 0.0 0.5763 0.0 0.5932 0.0 0.6102 0.0 0.6271 0.0 0.6441 0.0 0.661 0.0 0.678 0.0 0.6949 0.0 0.7119 0.0 0.7288 0.0 0.7458 0.0 0.7627 0.0 0.7797 0.0 0.7966 0.0 0.8136 0.0 0.8305 0.0 0.8475 0.0 0.8644 0.0 0.8814 0.0 0.8983 0.0 0.9153 0.0 0.9322 0.0 0.9492 0.0 0.9661 0.0 0.9831 0.0 1.0 0.0 /\sparkdot 0.163 1.0 black \end{sparkline} & 0.220000 \\
\bottomrule
\end{tabular}
}
      \subcaption{$MSE(\mathbf{C_{l_1}}+\mathbf{P_{l_1}})$ histogram}
      \label{tab:BMZI}
\end{minipage}

      \caption{BMZI summary results for QMIO quantum computer. (a) $\mathbf{C_{l_1}}+\mathbf{P_{l_1}}$ (black), $\mathbf{C_{l_1}}$ (orange), and $\mathbf{P_{l_1}}$ (blue) mean-value curves from the $m=128$ experiments for the $32$ QMIO qubits. The error bars are the standard deviations. The thin lines in the background are the theoretical curves for a pure state. (b) Histogram for the MSE of the sum of the coherence and the predictability generated from $m=128$ experiments. For each qubit, the table gathers the mean value (black middle dot) of the distribution in the range $[0,1]$ with its standard deviation, and the mininum (left black dot) and maximum (right black dot) values found for the MSE. It also shows the mean value of the correlation of the predictability and coherence with negative values painted in red. }
   \label{tab_fig:BMZI}
\end{figure*}

\section{RESULTS.}

\begin{figure*}[!ht]

\begin{minipage}[b]{0.45\textwidth}
   \centering
    \includegraphics[width=1\linewidth]{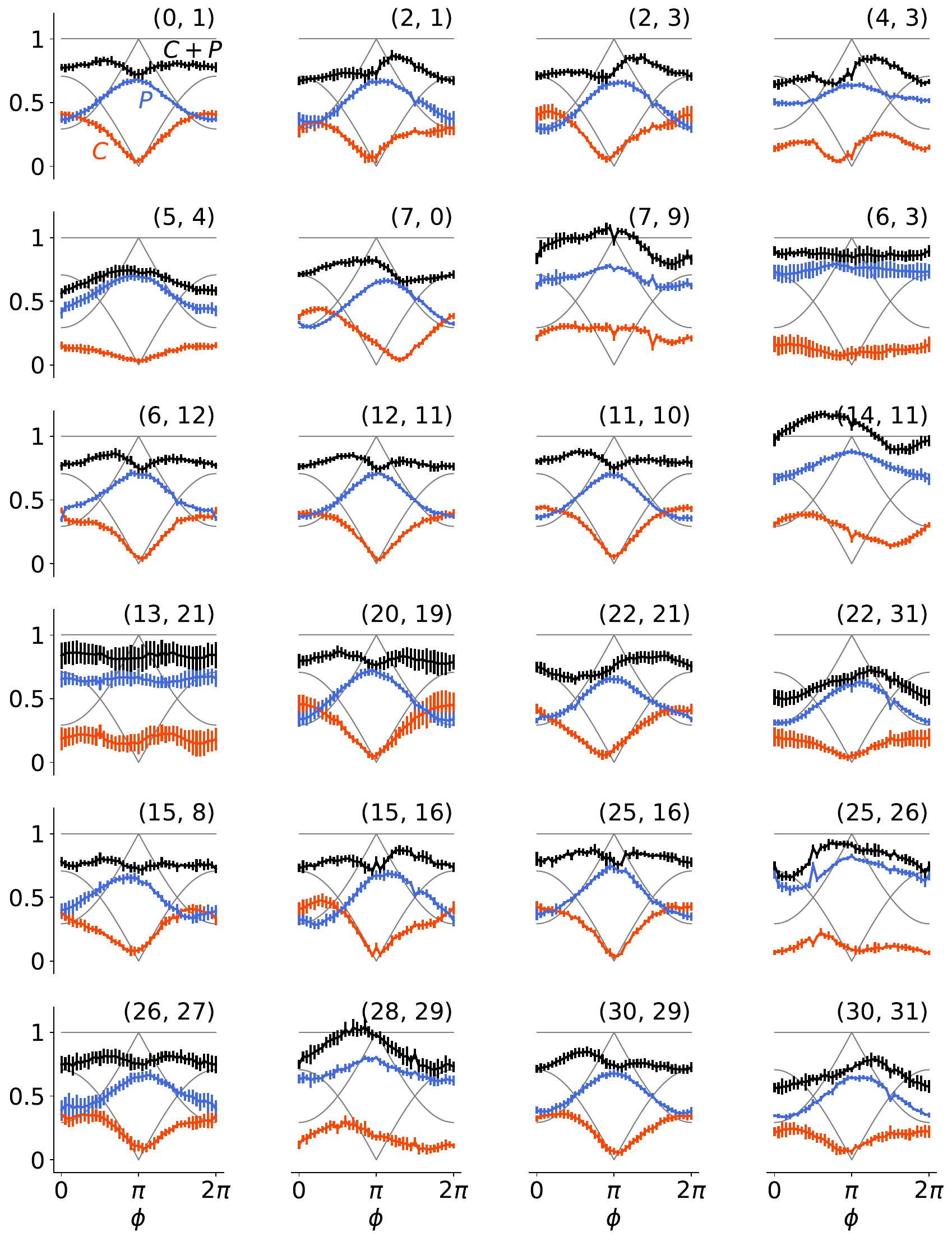}
      \bigskip
      \subcaption{BCP curves}  
      \label{fig:PQE}
\end{minipage} 
\hspace{0.5cm}  
\begin{minipage}[b]{0.45\textwidth}

\robustify\bfseries
\tabcolsep=0.11cm
\scalebox{0.7}{
\begin{tabular}{llS[table-format=2.5, round-mode=places,round-precision = 4]S[table-format=2.5, round-mode=places,round-precision = 4]S[table-format=2.5, round-mode=places,round-precision = 4]S[table-format=2.5, round-mode=places,round-precision = 4]cS[table-format=2.4, round-mode=places,round-precision = 3]}
\toprule
\textbf{qubit pair} & \textbf{m} & \textbf{mean} & \textbf{std} & \textbf{corr} & \textbf{min} & \textbf{MSE histogram} & \textbf{max} \\
\midrule
(0, 1) & 16 & 0.047200 & 0.003900 & -0.009300 & 0.038600 & \begin{sparkline}{15}\sparkdot 0.0538771239931769 0 black \sparkdot 0.03862504444527448 0 black \spark0.0 0.0 0.0169 0.0 0.0339 1.0 0.0508 0.2308 0.0678 0.0 0.0847 0.0 0.1017 0.0 0.1186 0.0 0.1356 0.0 0.1525 0.0 0.1695 0.0 0.1864 0.0 0.2034 0.0 0.2203 0.0 0.2373 0.0 0.2542 0.0 0.2712 0.0 0.2881 0.0 0.3051 0.0 0.322 0.0 0.339 0.0 0.3559 0.0 0.3729 0.0 0.3898 0.0 0.4068 0.0 0.4237 0.0 0.4407 0.0 0.4576 0.0 0.4746 0.0 0.4915 0.0 0.5085 0.0 0.5254 0.0 0.5424 0.0 0.5593 0.0 0.5763 0.0 0.5932 0.0 0.6102 0.0 0.6271 0.0 0.6441 0.0 0.661 0.0 0.678 0.0 0.6949 0.0 0.7119 0.0 0.7288 0.0 0.7458 0.0 0.7627 0.0 0.7797 0.0 0.7966 0.0 0.8136 0.0 0.8305 0.0 0.8475 0.0 0.8644 0.0 0.8814 0.0 0.8983 0.0 0.9153 0.0 0.9322 0.0 0.9492 0.0 0.9661 0.0 0.9831 0.0 1.0 0.0 /\sparkdot 0.0472 0.23076923076923078 black \end{sparkline} & 0.054000 \\
\bottomrule
\end{tabular}

}
      \subcaption{$MSE(\mathbf{C_{l_1}}+\mathbf{P_{l_1}})$ histogram}
      \label{tab:PQE}
\end{minipage}

      \caption{PQE summary results for QMIO quantum computer.(a) $\mathbf{C_{l_1}}+\mathbf{P_{l_1}}$ (black), $\mathbf{C_{l_1}}$ (orange), and $\mathbf{P_{l_1}}$ (blue) mean-value curves from the $m$ PQE experiments for the $32$ QMIO qubits. The error bars are the standard deviations. The thin lines in the background are the theoretical curves for a pure state. (b) Histogram for the MSE of the sum of the coherence and the predictability. For each qubit, the table gathers the mean value (black middle dot) of the distribution in the range $[0,1]$ with its standard deviation, and the miminum (left black dot) and maximum (right black dot) values found for the MSE. It also shows the correlation of the predictability and coherence with negative values painted in red. }
   \label{tab_fig:PQE}
\end{figure*}

 We executed the BMZI experiment from December 4, 2025, to January 6, 2026, and the PQE experiment from December 4, 2025, to January 2, 2026 on the 32-qubit quantum computer QMIO. Due to quantum state tomography, we require $3$ circuits for the BMZI experiment and $15$ for the PQE experiment to reconstruct the density matrix. We set $1000$ shots for each circuit. For details,  the algorithm repository is available in the CODE section.

We present the results as follows. First, we plot the mean value for $m$ BMZI/PQE experiments for the sum of coherence (quantified by the $l_1$ norm, $ \mathbf{C}_{l_1}$) and predictability ($ \mathbf{P}_{l_1}$), as $ \mathbf{C}_{l_1}+ \mathbf{P}_{l_1}$. We also show individual results for coherence and predictability. Second, we calculate the mean squared error (MSE) of the sum, $MSE(\mathbf{C}_{l_1}+ \mathbf{P}_{l_1})$. For $m$ BMZI and PQE experiments, we obtain $m$ values of this MSE and plot a histogram. We determine the mean and standard deviation of the histogram, and calculate the mean value of the $m$ correlations using equation (\ref{eq:corr}). We use sparklines—small, high-resolution graphics—to represent these histograms. Sparklines help visualize data patterns for statistical reflection, distinction, and comparison \cite{tufte2006beautiful}. 

We range the sparklines from $0$ to $1$. In a first approach, an experiment closer to the theoretical expectation reveals a left-shifted MSE-histogram sparkline. To unlock the hidden, we obtain the correlation between $\mathbf{C_{l_1}}$ and $\mathbf{P_{l_1}}$ to visualise its contribution to the $MSE(\mathbf{C_{l_1}}+\mathbf{P_{l_1}})$. Not showing the correlation is misinforming. Thus, more precisely: (1) a left-shifted MSE histogram with correlation tending to zero determines an experiment closer to the theoretical expectation, and (2) a narrow distribution relates to a reproducible experiment.

FIG. \ref{tab_fig:BMZI} encompasses the results for the BMZI experiment for all qubits available in QMIO. In FIG. \ref{fig:BMZI}, we can see the distinct mean curves from $128$ experiments for $\mathbf{C_{l_1}}+\mathbf{P_{l_1}}$, $\mathbf{C_{l_1}}$, and $\mathbf{P_{l_1}}$. At a glance, we see border effects in all of them, obtaining the absolute-extreme values for $\alpha=-\pi/2$. Moreover, we can see worse performances for qubits $4$, $13$, $31$, and $24$ with flat curves, and an asymmetrical performance for qubit $17$. 

Table in FIG. \ref{tab:BMZI} shows the associated sparklines for the $MSE(\mathbf{C_{l_1}}+\mathbf{P_{l_1}})$ histogram for the $m=128$ BMZI experiments.  We can see the expected deviations of the $\mathbf{C_{l_1}}+\mathbf{P_{l_1}}$-histogram to the right for qubits $4$, $13$, $31$, and $24$, as the plots show. Qubits $13$, $21$, $25$, exemplify wider distributions that imply less reproducibility of the BMZI experiment. This same information is revealed in the error bars of their curves. The first point related to the matching experiment-theory is more intricate. Moreover, qubit 17 is a clear example of a very sharp left-shifted distribution with a highly negative correlation. Not showing the curves or the correlation could lead us to misinterpretations of a good performance. This negative correlation drifts the values of $MSE(\mathbf{C_{l_1}}+\mathbf{P_{l_1}})$ to the left, faking the results: the MSE of the sum is low, but the MSE of coherence and predictability independently are high. Considering all the information gathered in the table, we could select qubit 20 as the one with the best performance.  

The results for the PQE experiment are summarised in FIG. \ref{tab_fig:PQE}. For this case, we select some pairs of qubits nearby and execute them in blocks of eight with $m\in[8,35]$. We can analyse the results from the MSE-histogram table in FIG.\ref{tab:PQE}. All $MSE(\mathbf{C_{l_1}}+\mathbf{P_{l_1}})$  histograms are left-shifted and sharp but they hide highly negative correlations for almost all pairs. Pairs $(0,1)$, $(2,3)$, $(11,10)$, $(20,19)$, $(22,21)$, $(25,16)$, $(30,31)$ display the lowest correlations. For this experiment, we also see values for $\mathbf{C_{l_1}}+\mathbf{P_{l_1}}$ in FIG. \ref{fig:PQE} higher than $1$ for the pairs $(14,11)$ and $(28,29)$  flat behaviours for the curves for pairs $(6,3)$ and $(13,21)$ with high correlations.\\

In this work, we prepared BMZI and PQE experiments using pure states that satisfy the equality in equation (\ref{eq:C+P}). Because these theoretical values of $C_{l_1}$ and $P_{l_1}$ are functionally dependent, we can analyse the mean squared error (MSE) of their sum with respect to the value of 1. Our results show that the experimental curves for $\mathbf{C_{l_1}}$, $\mathbf{P_{l_1}}$, and $\mathbf{C_{l_1}}+\mathbf{P_{l_1}}$ align with the theoretical predictions for pure states when both $MSE(\mathbf{C_{l_1}}+\mathbf{P_{l_1}})$ and the correlation between $\mathbf{C_{l_1}}$ and $\mathbf{P_{l_1}}$ defined in equation (\ref{eq:corr}) are low.

\section{CODE.}
The code with the packages to reproduce the executions is available in the repository: \begin{small}\url{https://gitlab.com/proyectos-cesga/quantum/pccc/bcp}.
\end{small}

\section{CONCLUSIONS.}

The complementarity relations are a quantitative formulation of Bohr's Complementarity Principle. Certain experimental setups, such as the ones presented in this work, enable the study of the wave-particle duality by allowing a partial manifestation of both these complementary aspects of quantum mechanics.
We present theoretical implementations of the Biased Mach-Zehnder interferometer and the Partial Quantum Eraser as one and two-qubit circuits leading to pure final states. Multiple executions were carried out on the 32-qubit quantum computer QMIO from December 4 to January 6 and from December 4 to January 2 for each of the mentioned experiments, respectively. In practice, to reconstruct the final density matrix, we use quantum state tomography to assess coherence and predictability for each experiment. Mean values for coherence, predictability and their sum are graphically presented and compared to the expected theoretical outcomes and a mean squared error (MSE) analysis is conducted for the sum of coherence and predictability. From the collation of both the graphical results and the MSE analysis, the following can be inferred: a value close to zero for the MSE alone can mislead one into interpreting the performance of the experiment as good when it may be hiding a significant contribution of the correlation between coherence and predictability. Low MSE and correlation evidence a final state close to being pure.\\

This work broadens the analysis of the purity of qubit states. In terms of methodology, delving into another statistical qubit analysis, such as ANOVA, could establish levels of performance or even comparisons between qubits.  Furthermore, we could benchmark different quantum platforms to compare  device performance. To conclude, these interferometer tests on a real quantum computer contribute to the open debate on interpreting the nature of quantum theory.

\begin{acknowledgements}


This article shows the results from the executions of two experiments to test Bohr's Complementarity Principle on a real Quantum Computer. The initial development stems from the work carried out in a Physics Bachelor's one month research traineeship at Galicia Supercomputing Center (CESGA) and later on continued and executed on CESGA's  Quantum Computer QMIO. 
Co-author Celia Álvarez Álvarez (a bachelor's student at the time of publication) would like to thank CESGA for providing the opportunity to expand upon the initial work. She also extends special thanks to co-author Mariamo Mussa Juane for all the valuable discussions and time generously devoted to them.\\

We specially thank Francisco Javier Cardama for the creation of the quantum-circuit images. We also thank Álvaro C. Tabarés for the automatisation of the executions of both BMZI and PQE experiments. Finally, we thank Francisco Yáñez Rodríguez for the thinking-out-of-the-box discussion on the MSE analysis.  \\

We thank the CESGA and specially the Quantum Computing group members for their technical assistance, feedback, and the stimulating intellectual environment they provide. This work was supported by Axencia Galega de Innovación through the Grant Agreement “Despregamento dunha infraestrutura baseada en tecnoloxías cuánticas da información que permita impulsar a I+D+i en Galicia” within the program FEDER Galicia 2014-2020. Mariamo Mussa Juane was supported by MICIN through the European Union NextGenerationEU recovery plan (PRTR-C17.I1), and by the Galician Regional Government through the “Planes Complementarios de I+D+i con las Comunidades Autónomas” in Quantum Communication. Simulations are performed using the Finisterrae III Supercomputer, funded by the project CESGA-01 FINISTERRAE III. QMIO infrastructure is funded by the European Union, through the Programa Operativo Galicia 2014-2020 of ERDF$\_$REACT EU, as  part of the European Union's response to the COVID-19 pandemic.

\end{acknowledgements}

\newpage

\nocite{*}

\bibliography{apssamp}

\end{document}